\documentclass[aps,prl,twocolumn,superscriptaddress]{revtex4}
\usepackage{amssymb,amsbsy,graphicx,xcolor,times,subfigure,marvosym,color,bm,multirow,epstopdf}
\usepackage[latin1]{inputenc}
\begin{document}

\title{Fracton topological phases from strongly coupled spin chains}

\author{G\'abor B. Hal\'asz}
\affiliation{Kavli Institute for Theoretical Physics, University of
California, Santa Barbara, CA 93106, USA}

\author{Timothy H. Hsieh}
\affiliation{Kavli Institute for Theoretical Physics, University of
California, Santa Barbara, CA 93106, USA}

\author{Leon Balents}
\affiliation{Kavli Institute for Theoretical Physics, University of
California, Santa Barbara, CA 93106, USA}


\begin{abstract}

We provide a new perspective on fracton topological phases, a class
of three-dimensional topologically ordered phases with
unconventional fractionalized excitations that are either completely
immobile or only mobile along particular lines or planes. We
demonstrate that a wide range of these fracton phases can be
constructed by strongly coupling mutually intersecting spin chains
and explain via a concrete example how such a coupled-spin-chain
construction illuminates the generic properties of a fracton phase.
In particular, we describe a systematic translation from each
coupled-spin-chain construction into a parton construction where the
partons correspond to the excitations that are mobile along lines.
Remarkably, our construction of fracton phases is inherently based
on spin models involving only two-spin interactions and thus brings
us closer to their experimental realization.

\end{abstract}


\maketitle


One of the most striking features of topologically ordered phases in
two dimensions is the existence of quasiparticle excitations with
fractional quantum numbers and fractional exchange statistics
\cite{Kitaev-2006}. In three dimensions, this fractionalization
attains an even more exotic character and has proven to be a vast
and exciting frontier. For example, there are loop-like excitations
in addition to point-like excitations, and the intricate braiding
patterns exhibited by these loop-like excitations are essential for
characterizing the topological order \cite{Wang-2014, Jiang-2014}.

Fracton topological phases are topologically ordered phases in three
dimensions with a particularly extreme form of fractionalization
\cite{Chamon-2005, Bravyi-2011a, Haah-2011, Yoshida-2013,
Vijay-2015, Pretko-2017}. In these phases, there are point-like
excitations that are either completely immobile or only mobile in a
lower-dimensional subsystem, such as an appropriate line or plane.
Remarkably, the restricted mobility of excitations has a purely
topological origin and appears in translation-invariant systems
without any disorder. In addition to being of fundamental interest
from the perspective of topological phases, and providing an
exciting disorder-free alternative to many-body localization
\cite{Kim-2016, Prem-2017}, this phenomenology has important
implications for quantum-information storage. Indeed, the immobility
of excitations makes encoded quantum information more stable at
finite temperature than in conventional topologically ordered phases
\cite{Bravyi-2011b, Bravyi-2013}.

In recent years, several different viewpoints have been presented on
fracton topological phases. From a purely conceptual perspective,
fracton phases can be understood by gauging classical spin models
with particular subsystem symmetries \cite{Vijay-2016,
Williamson-2016} or in terms of generalized parton constructions
with overlapping directional gauge constraints and/or interacting
parton Hamiltonians \cite{Hsieh-2017}. While these approaches can be
used to understand the generic properties of fracton phases, the
concrete spin models they provide are far from realistic as they
involve interactions between many spins at the same time. From a
more practical perspective, fracton phases can be constructed by
coupling orthogonal stacks of two-dimensional topologically ordered
layers \cite{Ma-2017, Vijay-2017}. This approach can lead to more
realistic spin models involving only two-spin interactions
\cite{Slagle-2017}, although it is not immediately clear what kind
of fracton phase is obtained from a generic construction.

In this Letter, we provide an understanding of fracton topological
phases in terms of coupled spin chains and, along with it, a
systematic route to construct realistic spin models hosting such
fracton phases. This coupled-spin-chain construction is useful for
three main reasons. First, like all coupled-chain (i.e.,
coupled-wire) constructions, it decomposes the system into its most
basic building blocks, and dealing directly with these building
blocks offers significant versatility in describing a rich variety
of fracton phases. Second, the coupled-spin-chain constructions
directly translate into generalized parton constructions, and the
generic properties of the corresponding fracton phases can then be
readily understood. For example, one can immediately identify the
excitations with restricted mobility and their respective
lower-dimensional subsystems (i.e., lines or planes). Third, the
coupled-spin-chain constructions naturally give rise to fracton spin
models involving only two-spin interactions, which are more amenable
to a potential experimental implementation.

\begin{figure}
\includegraphics[width=0.98\columnwidth]{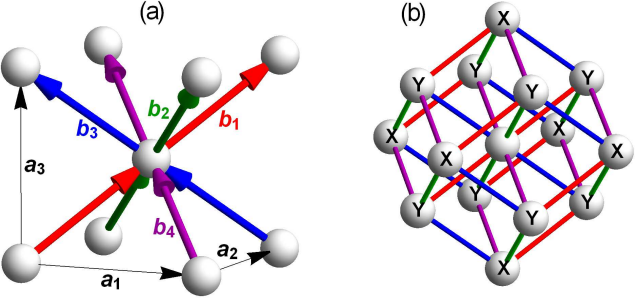}
\caption{Fracton spin model on the BCC lattice characterized by
cubic lattice vectors $\mathbf{a}_{1,2,3}$ and bond vectors
$\mathbf{b}_{1,2,3,4}$. (a) Nearest-neighbor terms of the model
Hamiltonian $H$. Each term corresponding to a $j = 1$ (red), $j = 2$
(green), $j = 3$ (blue), or $j = 4$ (purple) bond acts on spin
flavor $j$ via $\sigma^x$ at the tail and via $\sigma^y$ at the head
of the bond arrow. (b) Effective Hamiltonian $\tilde{H}$ in the
strong-coupling limit. Each term $W_{\mathbf{r}}$ in $\tilde{H}$ is
induced by nearest-neighbor terms (colored lines) in degenerate
perturbation theory and is a product of eight spin operators
$Y_{\tilde{\mathbf{r}}}$ at the corners of the basic BCC cube as
well as six spin operators $X_{\tilde{\mathbf{r}}}$ at the apices of
the square pyramids based on the faces of this cube. All sites
$\tilde{\mathbf{r}}$ are marked by appropriate labels.}
\label{fig-1}
\end{figure}

\emph{Fracton spin model.---}Our coupled-spin-chain construction
works for any $4n$-coordinated ($n \geq 2$) lattice with $2n$
spin-one-half degrees of freedom per site. For concreteness,
however, we concentrate on the eight-coordinated ($n = 2$)
body-centered-cubic (BCC) lattice, which is characterized by the
(conventional) cubic lattice vectors $\mathbf{a}_{1,2,3}$ and the
nearest-neighbor bond vectors $\mathbf{b}_{1,2,3,4}$ [see
Fig.~\ref{fig-1}(a)].

In the concrete model, there are four spins $\sigma_{\mathbf{r}, j}$
with flavors $j = 1,2,3,4$ at each site $\mathbf{r}$ of the BCC
lattice, and the Hamiltonian in terms of these spins reads
\begin{equation}
H = -J \sum_j \sum_{\langle \mathbf{r}, \mathbf{r}' \rangle_j}
\sigma_{\mathbf{r}, j}^x \sigma_{\mathbf{r}', j}^y - \lambda J
\sum_{\mathbf{r}} \sum_{\langle j,j' \rangle} \sigma_{\mathbf{r},
j}^z \sigma_{\mathbf{r}, j'}^z, \label{eq-H-1}
\end{equation}
where $\langle j,j' \rangle$ implies a summation over all pairs of
spins at the same site, and $\langle \mathbf{r}, \mathbf{r}'
\rangle_j$ implies a summation over all $j$ bonds ($j = 1,2,3,4$)
such that the arrow in Fig.~\ref{fig-1}(a) points from $\mathbf{r}$
to $\mathbf{r}'$ at each bond. The first (nearest-neighbor) term
describes decoupled spin chains of the four spin flavors along the
$\langle 1 \, 1 \, 1 \rangle$ directions traced out by strings of
the four corresponding bond types, while the second (on-site) term
introduces a coupling between spin chains of distinct spin flavors
intersecting at any site. Note that the individual (decoupled) spin
chains are both critical and macroscopically degenerate.

In the strong-coupling regime ($\lambda \gg 1$), the four spins
$\sigma_{\mathbf{r}, j}$ at each site $\mathbf{r}$ are locked
together by the on-site terms, and thus $\sigma_{\mathbf{r}, j}^z =
\sigma_{\mathbf{r}, j'}^z$ for all $j$ and $j'$. The local Hilbert
space is then captured by a single effective spin
$\Sigma_{\mathbf{r}}$ as its two states can be characterized by
$\Sigma_{\mathbf{r}}^z = \sigma_{\mathbf{r}, j}^z = \pm 1$. For
$\lambda \rightarrow \infty$, these degenerate local states give
rise to an exponentially large ground-state degeneracy. However, if
$\lambda$ is finite, the nearest-neighbor terms select particular
superpositions of these ground states by inducing a low-energy
Hamiltonian within the ground-state subspace in terms of the
effective spin components
\begin{eqnarray}
X_{\mathbf{r}} &\equiv& \Sigma_{\mathbf{r}}^x = \sigma_{\mathbf{r},
1}^x \sigma_{\mathbf{r}, 2}^x \sigma_{\mathbf{r}, 3}^x
\sigma_{\mathbf{r}, 4}^x = -\sigma_{\mathbf{r}, 1}^y
\sigma_{\mathbf{r}, 2}^y \sigma_{\mathbf{r}, 3}^x
\sigma_{\mathbf{r}, 4}^x = \ldots,
\nonumber \\
Y_{\mathbf{r}} &\equiv& \Sigma_{\mathbf{r}}^y = \sigma_{\mathbf{r},
1}^y \sigma_{\mathbf{r}, 2}^x \sigma_{\mathbf{r}, 3}^x
\sigma_{\mathbf{r}, 4}^x = \sigma_{\mathbf{r}, 1}^x
\sigma_{\mathbf{r}, 2}^y \sigma_{\mathbf{r}, 3}^x
\sigma_{\mathbf{r}, 4}^x = \ldots,
\nonumber \\
Z_{\mathbf{r}} &\equiv& \Sigma_{\mathbf{r}}^z = \sigma_{\mathbf{r},
1}^z = \sigma_{\mathbf{r}, 2}^z = \sigma_{\mathbf{r}, 3}^z =
\sigma_{\mathbf{r}, 4}^z. \label{eq-XYZ-1}
\end{eqnarray}
For our BCC model in Eq.~(\ref{eq-H-1}), the lowest-order
non-trivial Hamiltonian term $W_{\mathbf{r}}$ arises at order $32$
in degenerate perturbation theory (see the Supplementary Material
\cite{Supplement}) and is a product of $14$ effective spin operators
[see Fig.~\ref{fig-1}(b)]. Ignoring any trivial (i.e., constant)
terms, the effective Hamiltonian at this order is then $\tilde{H} =
\sum_{\mathbf{r}} W_{\mathbf{r}}$, where
\begin{equation}
W_{\mathbf{r}} \sim \frac{J} {\lambda^{31}} \prod_{\pm}
X_{\mathbf{r} \pm \mathbf{a}_1} X_{\mathbf{r} \pm \mathbf{a}_2}
X_{\mathbf{r} \pm \mathbf{a}_3} Y_{\mathbf{r} \pm \mathbf{b}_1}
Y_{\mathbf{r} \pm \mathbf{b}_2} Y_{\mathbf{r} \pm \mathbf{b}_3}
Y_{\mathbf{r} \pm \mathbf{b}_4}. \label{eq-W}
\end{equation}
Since $[W_{\mathbf{r}}, W_{\mathbf{r}'}] = 0$ for all $\mathbf{r}$
and $\mathbf{r}'$, the Hamiltonian $\tilde{H}$ corresponds to a
commuting-projector model, where each eigenstate is characterized by
$W_{\mathbf{r}} = \pm 1$. Furthermore, the only non-trivial terms
arising at higher orders of perturbation theory are products of
$W_{\mathbf{r}}$, and this commuting-projector model thus captures
an entire strong-coupling phase $\lambda > \lambda_C$ above a
critical coupling strength $\lambda_C$.

\begin{figure}
\includegraphics[width=0.98\columnwidth]{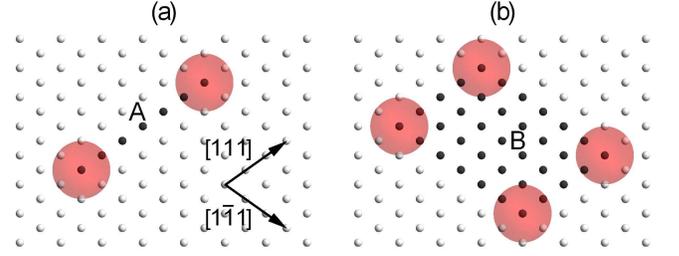}
\caption{One-dimensional (a) and zero-dimensional (b) excitations in
the $(1 \, 0 \, \bar{1})$ plane of our BCC model. In each case, the
excitations are (schematically) located within the red circles and
are created by the operator $\prod_{\mathbf{r} \in A, B}
Z_{\mathbf{r}}$ over the sites $\mathbf{r} \in A, B$ marked by black
dots.} \label{fig-2}
\end{figure}

This strong-coupling phase of the model in Eq.~(\ref{eq-H-1}) is
identified as a type-I fracton phase \cite{Vijay-2016}, which is
characterized by the following (closely related) features. First of
all, there is a ground-state degeneracy that scales as $\sim 2^L$
with the linear system dimension $L$ due to the planar conservation
laws $\prod_{\mathbf{r} \in \{ 1 \, 1 \, 0 \}} W_{\mathbf{r}} =
\mathrm{const.}$ within the $\{ 1 \, 1 \, 0 \}$ planes of the
lattice \cite{Footnote-1}. For a product $\prod_{\mathbf{r} \in R
\subset \{ 1 \, 1 \, 0 \}} W_{\mathbf{r}}$ within a finite region
$R$ of a $\{ 1 \, 1 \, 0 \}$ plane, the boundary of the region then
corresponds to a string logical operator, and the excitations at the
endpoints of such a string $\partial R$ are only mobile within the
given $\{ 1 \, 1 \, 0 \}$ plane. Moreover, there is a string logical
operator $\prod_{\mathbf{r} \in A} Z_{\mathbf{r}}$ along each
$\langle 1 \, 1 \, 1 \rangle$ direction of the lattice, and the
excitations at the endpoints of such a string $A$ are only mobile
along the given $\langle 1 \, 1 \, 1 \rangle$ direction [see
Fig.~\ref{fig-2}(a)]. Finally, these strings can be assembled into
membrane logical operators $\prod_{\mathbf{r} \in B} Z_{\mathbf{r}}$
within parallelepipeds spanned by two distinct $\langle 1 \, 1 \, 1
\rangle$ directions (e.g., the $[1 \, 1 \, 1]$ and the $[1 \,
\bar{1} \, 1]$ directions), and the excitations at the corners of
such a parallelepiped $B$ are completely immobile [see
Fig.~\ref{fig-2}(b)].

\emph{Parton decomposition.---}The effective spin Hamiltonian
$\tilde{H}$ has an exact non-interacting parton construction.
Indeed, the spins $\Sigma_{\mathbf{r}}$ can be decomposed into
clusters of partons that are individually governed by a
non-interacting Hamiltonian but are also subject to gauge
constraints that recombine them into their parent spins. Such parton
constructions are commonly used to capture strongly correlated spin
phases, including spin liquids, on a variational level
\cite{Wen-2002}.

\begin{figure}
\includegraphics[width=0.98\columnwidth]{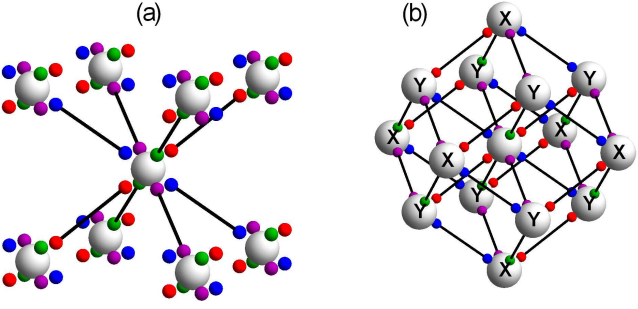}
\caption{Exact parton construction of our BCC model. (a) Each spin
$\Sigma_{\mathbf{r}}$ (large white sphere) is decomposed into eight
Majorana fermions (colored dots) at the bonds around the site
$\mathbf{r}$: four $\gamma_{\mathbf{r}, j}$ above $\mathbf{r}$ and
four $\hat{\gamma}_{\mathbf{r}, j}$ below $\mathbf{r}$ with flavors
$j = 1$ (red), $j = 2$ (green), $j = 3$ (blue), and $j = 4$
(purple). Each bond is occupied by two Majorana fermions
$\gamma_{\mathbf{r}, j}$ and $\hat{\gamma}_{\mathbf{r}', j}$ that
are in a state characterized by the bond-fermion operator $i
\gamma_{\mathbf{r}, j} \hat{\gamma}_{\mathbf{r}', j} = \pm 1$ (black
line). (b) Decomposition of each term $W_{\mathbf{r}}$ in the
effective spin Hamiltonian $\tilde{H}$ [see Fig.~\ref{fig-1}(b)]
into a product of bond-fermion operators (black lines).}
\label{fig-3}
\end{figure}

For the eight-coordinated BCC lattice, it is a natural choice
\cite{Wen-2003} to decompose each spin $\Sigma_{\mathbf{r}}$ into
eight Majorana fermions (partons) $\gamma_{\mathbf{r}, j}$ and
$\hat{\gamma}_{\mathbf{r}, j}$ with flavors $j = 1,2,3,4$ and to
assign these eight partons to the eight respective bonds around the
site $\mathbf{r}$ [see Fig.~\ref{fig-3}(a)]. The two Majorana
fermions at each bond then form a complex fermion, which is demanded
to be in an occupied or an unoccupied state, and the parton state is
simply the direct product of all these local states. Formally, the
parton state is the ground state of the non-interacting Hamiltonian
\begin{equation}
\mathcal{H} = \sum_j \sum_{\langle \mathbf{r}, \mathbf{r}'
\rangle_j} i \nu_{\mathbf{r}, \mathbf{r}'} \gamma_{\mathbf{r}, j}
\hat{\gamma}_{\mathbf{r}', j}, \label{eq-H-2}
\end{equation}
where $\nu_{\mathbf{r}, \mathbf{r}'} = \pm 1$ determines whether the
complex fermion at the bond $\langle \mathbf{r}, \mathbf{r}'
\rangle_j$ is occupied or unoccupied.

Since the parton decomposition increases the local Hilbert space at
each site, the partons must be reconciled with their parent spins by
means of appropriate gauge constraints. Following
Ref.~\cite{Hsieh-2017}, we capture our type-I fracton phase by
imposing the overlapping directional gauge constraints
\begin{equation}
G_{\mathbf{r}, j, j'} = \gamma_{\mathbf{r}, j} \gamma_{\mathbf{r},
j'} \hat{\gamma}_{\mathbf{r}, j} \hat{\gamma}_{\mathbf{r}, j'} = 1.
\label{eq-G}
\end{equation}
These gauge constraints are indeed directional as each of them only
acts on partons in a particular $\{ 1 \, 1 \, 0 \}$ plane and
overlapping as any two such planes intersect along a particular
$\langle 1 \, 1 \, 1 \rangle$ direction. We also note that there are
three independent gauge constraints at each site which correctly
reconcile eight Majorana fermions with a single spin.

The three components of the spin $\Sigma_{\mathbf{r}}$ are
identified with the three inequivalent gauge-invariant operators
\begin{eqnarray}
X_{\mathbf{r}} &=& \gamma_{\mathbf{r}, 1} \gamma_{\mathbf{r}, 2}
\gamma_{\mathbf{r}, 3} \gamma_{\mathbf{r}, 4} =
-\hat{\gamma}_{\mathbf{r}, 1} \hat{\gamma}_{\mathbf{r}, 2}
\gamma_{\mathbf{r}, 3} \gamma_{\mathbf{r}, 4} = \ldots,
\nonumber \\
Y_{\mathbf{r}} &=& \hat{\gamma}_{\mathbf{r}, 1} \gamma_{\mathbf{r},
2} \gamma_{\mathbf{r}, 3} \gamma_{\mathbf{r}, 4} =
\gamma_{\mathbf{r}, 1} \hat{\gamma}_{\mathbf{r}, 2}
\gamma_{\mathbf{r}, 3} \gamma_{\mathbf{r}, 4} = \ldots,
\label{eq-XYZ-2} \\
Z_{\mathbf{r}} &=& i \hat{\gamma}_{\mathbf{r}, 1}
\gamma_{\mathbf{r}, 1} = i \hat{\gamma}_{\mathbf{r}, 2}
\gamma_{\mathbf{r}, 2} = i \hat{\gamma}_{\mathbf{r}, 3}
\gamma_{\mathbf{r}, 3} = i \hat{\gamma}_{\mathbf{r}, 4}
\gamma_{\mathbf{r}, 4}, \nonumber
\end{eqnarray}
where the equivalent expressions of each spin component are related
by the gauge constraints $G_{\mathbf{r}, j, j'}$. Each term
$W_{\mathbf{r}}$ in the spin Hamiltonian $\tilde{H}$ is then readily
written in terms of the partons and decomposes into a product of
$32$ bond-fermion operators $i \gamma_{\mathbf{r}, j}
\hat{\gamma}_{\mathbf{r}', j}$ in Eq.~(\ref{eq-H-2}) [see
Fig.~\ref{fig-3}(b)]. Since the terms $W_{\mathbf{r}}$ also commute
with the gauge constraints, the exact eigenstates of the spin
Hamiltonian $\tilde{H}$ are thus obtained from those of the
(non-interacting) parton Hamiltonian $\mathcal{H}$ by enforcing the
gauge constraints via appropriate projections.

From a comparison of Figs.~\ref{fig-1} and \ref{fig-3}, there is
clearly an intimate connection between the coupled-spin-chain
construction in Eq.~(\ref{eq-H-1}) and the parton construction in
Eq.~(\ref{eq-H-2}). Indeed, the spin-combination rules in
Eq.~(\ref{eq-XYZ-1}) for obtaining the effective low-energy
Hamiltonian $\tilde{H}$ in degenerate perturbation theory are
identical to the corresponding parton-decomposition rules in
Eq.~(\ref{eq-XYZ-2}) via the substitutions $\sigma_{\mathbf{r}, j}^x
\leftrightarrow \gamma_{\mathbf{r}, j}^{\phantom{*}}$,
$\sigma_{\mathbf{r}, j}^y \leftrightarrow \hat{\gamma}_{\mathbf{r},
j}^{\phantom{*}}$, and $\sigma_{\mathbf{r}, j}^z \leftrightarrow i
\hat{\gamma}_{\mathbf{r}, j}^{\phantom{*}} \gamma_{\mathbf{r},
j}^{\phantom{*}}$. This connection can be understood by means of the
Jordan-Wigner transformation
\begin{eqnarray}
\left( \begin{array}{c} \sigma_{\mathbf{r}, j}^{x} \\
\sigma_{\mathbf{r}, j}^{y}
\end{array} \right) = \Bigg[ \prod_{\mathbf{r}' <
\mathbf{r}} \prod_{j'} \sigma_{\mathbf{r}', j'}^z \prod_{j' < j}
\sigma_{\mathbf{r}, j'}^z \left( i \sigma_{\mathbf{r}, j}^z
\right)^{j-1} \Bigg] \left( \begin{array}{c} \gamma_{\mathbf{r},
j}^{\phantom{*}} \\ \hat{\gamma}_{\mathbf{r}, j}^{\phantom{*}}
\end{array} \right), \nonumber
\end{eqnarray}
\begin{equation}
\sigma_{\mathbf{r}, j}^z = i \hat{\gamma}_{\mathbf{r},
j}^{\phantom{*}} \gamma_{\mathbf{r}, j}^{\phantom{*}}, \label{eq-JW}
\end{equation}
where the additional factor $(i \sigma_{\mathbf{r}, j}^z)^{j-1}$
with respect to the standard form is a local spin rotation. Within
the low-energy subspace characterized by $\Sigma_{\mathbf{r}}$, the
Jordan-Wigner strings in the square brackets then disappear due to
the spin-locking constraints $\sigma_{\mathbf{r}, j}^z
\sigma_{\mathbf{r}, j'}^z = 1$ or, equivalently, due to the
corresponding gauge constraints $G_{\mathbf{r}, j, j'} = 1$. We
emphasize, however, that this connection is restricted to the
low-energy subspace and that it would thus be incorrect to argue for
Eq.~(\ref{eq-H-2}) by directly substituting Eq.~(\ref{eq-JW}) into
Eq.~(\ref{eq-H-1}).

\emph{Extended fracton phase.---}As discussed in
Ref.~\cite{Hsieh-2017}, parton constructions can be used to
understand the generic properties of fracton phases. In general, a
strongly correlated spin phase is characterized by its parton
construction via the invariant gauge group (IGG), which consists of
all gauge transformations (i.e., generic products of local gauge
constraints) that commute with the parton Hamiltonian.

For a type-I fracton phase, the IGG is generically $\mathbb{Z}_2^N$
with $N \sim L$ due to the presence of planar IGG elements that are
related to the planar conservation laws of the corresponding spin
model \cite{Hsieh-2017}. For our BCC construction, in particular,
there is a planar IGG element for each $\{ 1 \, 1 \, 0 \}$ plane as
the product of all $G_{\mathbf{r}, j, j'}$ in a $\{ 1 \, 1 \, 0 \}$
plane spanned by a net of $j$ and $j'$ bonds commutes with
$\mathcal{H}$ in Eq.~(\ref{eq-H-2}). The partons themselves can then
be identified with the excitations that are only mobile along
particular $\langle 1 \, 1 \, 1 \rangle$ directions. Indeed, since
each parton $\gamma_{\mathbf{r}, j}$ ($\hat{\gamma}_{\mathbf{r},
j}$) anticommutes with three planar IGG elements containing
$G_{\mathbf{r}, j, j'}$ with $j' \neq j$, it is constrained to move
along the intersection of the three corresponding planes, which is a
$\langle 1 \, 1 \, 1 \rangle$ direction traced out by a string of
$j$ bonds.

\begin{figure}
\includegraphics[width=0.88\columnwidth]{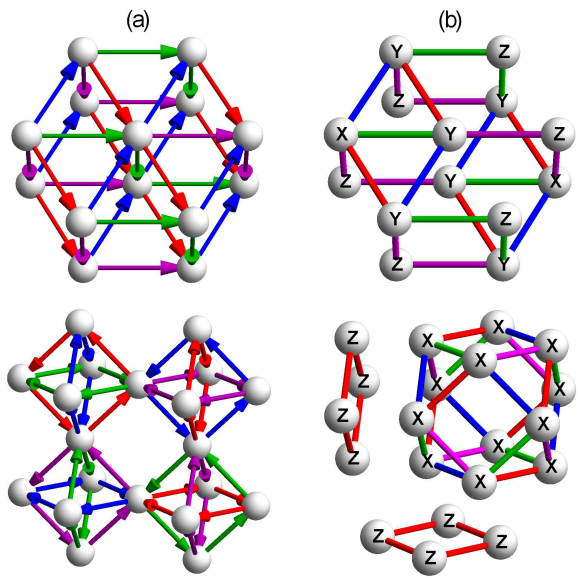}
\caption{Coupled-spin-chain constructions (a) and effective
strong-coupling Hamiltonians (b) capturing type-I fracton phases on
two different lattices. The notation is taken from Fig.~\ref{fig-1}.
For the second construction, the strong-coupling Hamiltonian has
three independent terms and corresponds to the X-cube model
\cite{Vijay-2016}.} \label{fig-4}
\end{figure}

Importantly, the parton construction is valid beyond the exactly
solvable model $\tilde{H}$. In fact, any sufficiently weak local
perturbation that commutes with all the IGG elements can be added to
Eq.~(\ref{eq-H-2}) while leaving the projected parton ground state
in the original fracton phase. In addition to the terms $i
\gamma_{\mathbf{r}, j} \hat{\gamma}_{\mathbf{r} + \mathbf{b}_j, j}$
already present, the generic quadratic terms appearing are then $i
\tilde{\gamma}_{\mathbf{r}, j} \tilde{\gamma}_{\mathbf{r} + x
\mathbf{b}_j, j}$, where $x$ is an arbitrary integer, and
$\tilde{\gamma}_{\mathbf{r}, j}$ is either $\gamma_{\mathbf{r}, j}$
or $\hat{\gamma}_{\mathbf{r}, j}$. In turn, these generic terms lead
to non-trivial parton dispersions along the respective $\langle 1 \,
1 \, 1 \rangle$ directions of motion. While the resulting parton
ground state does not correspond to an exactly solvable spin model,
it can be used as the starting point of a variational description.

\emph{Generalized constructions.---}Our coupled-spin-chain
construction is extremely versatile and readily generalizes to a
rich variety of fracton phases. First, it can be defined on any
$4n$-coordinated ($n \geq 2$) lattice with $2n$ spins
$\sigma_{\mathbf{r}, j = 1, \ldots, 2n}$ at each site $\mathbf{r}$.
Second, the intersecting spin chains can be embedded in the lattice
in many different ways. In particular, they do not have to follow
straight lines and might even connect back into themselves to form
closed loops.

Formally, the Hamiltonian is Eq.~(\ref{eq-H-1}) for any such
coupled-spin-chain construction, where the different bond types are
assigned to the given lattice in a particular way. It is crucial
that there are precisely two bonds of each type $j$ around each site
$\mathbf{r}$ at which the two corresponding terms act with spin
operators $\sigma_{\mathbf{r}, j}^x$ and $\sigma_{\mathbf{r}, j}^y$,
respectively. Two examples of such generalized constructions are
presented in Fig.~\ref{fig-4}(a) on a primitive hexagonal lattice
and on a cubic lattice formed by corner-sharing octahedra. For each
construction, there is a type-I fracton phase in the strong-coupling
limit, and the independent terms $W_{\mathbf{r}}$ of the effective
strong-coupling Hamiltonian $\tilde{H}$ are given in
Fig.~\ref{fig-4}(b). Remarkably, the type-I fracton phase of the
second construction is captured by the X-cube model
\cite{Vijay-2016}.

Moreover, the fracton phase in the strong-coupling limit can be
readily analyzed without obtaining the concrete form of the
effective Hamiltonian $\tilde{H}$. Due to the connection between the
coupled-spin-chain construction and the parton construction, the
terms $W_{\mathbf{r}}$ in $\tilde{H}$ necessarily decompose into
products of appropriate bond-fermion operators when written in terms
of the partons. The fracton phase is then captured by the
non-interacting parton Hamiltonian in Eq.~(\ref{eq-H-2}), where the
different bond types are assigned to the lattice in the same way as
in Eq.~(\ref{eq-H-1}). For such a generalized parton construction,
the IGG elements are products of gauge constraints $G_{\mathbf{r},
j, j'}$ along nets of $j$ and $j'$ bonds, while the partons
themselves correspond to excitations that are mobile along
respective strings of $j$ bonds (i.e., the individual spin chains).

\emph{Summary and outlook.---}We have provided a general framework
for describing fracton topological phases in terms of an
interpenetrating set of spin chains that are strongly coupled at
their intersection points. It is clear from the examples presented
that this construction can easily describe many different fracton
phases by spin models involving only two-spin interactions. This
work covers the strong-coupling limit of these spin models, while
the weak-coupling limit and the quantitative domain of the fracton
phase in the strong-coupling regime (i.e., the value of $\lambda_C$)
remain to be understood.

Our construction of fracton phases is analogous to how the
toric-code model is obtained in the spatially anisotropic limit of
the Kitaev honeycomb model \cite{Kitaev-2006}. Indeed, if we form
two pairs out of the four spin flavors in Eq.~(\ref{eq-H-1}) and
only introduce couplings within each pair, we obtain two orthogonal
stacks of two-dimensional topologically ordered layers; see the
Supplementary Material \cite{Supplement}. The fracton phase is then
recovered by including the remaining couplings between the two
orthogonal stacks \cite{Slagle-2017}. In a conceptual sense, the
coupled-layer models of fracton phases introduced in
Refs.~\cite{Ma-2017} and \cite{Vijay-2017} are thus an intermediate
step between our coupled-spin-chain models and the
commuting-projector models in Ref.~\cite{Vijay-2016}.

Finally, it follows from our work that parton constructions
describing fracton phases can be generally converted into
appropriate spin models. While the non-interacting parton
constructions in this work give rise to coupled-spin-chain models
involving two-spin interactions, the interacting parton
constructions in Ref.~\cite{Hsieh-2017} translate into more general
spin models involving four-spin interactions. Remarkably, these
parton constructions describe both type-I and type-II fracton
phases, characterized by immobile excitations at the corners of
membrane and fractal operators, respectively \cite{Footnote-2}. Our
formalism thus brings us one step closer to realizing these highly
unconventional topological phases in the laboratory.

G.~B.~H. and T.~H.~H. are supported by the Gordon and Betty Moore
Foundation's EPiQS Initiative through Grant No.~GBMF4304. L.~B. was
supported by the National Science Foundation under Grant No.~NSF
DMR1506119.



\clearpage

\begin{widetext}

\subsection{\large Supplementary Material}

\section{Degenerate perturbation theory} \label{sec-pert}

Here we describe the degenerate perturbation theory in the
strong-coupling limit ($\lambda \gg 1$) of our coupled-spin-chain
Hamiltonian $H$ and explain how the lowest-order effective
Hamiltonian $\tilde{H}$ is obtained. From the perspective of
perturbation theory, the coupled-spin-chain Hamiltonian in Eq.~(1)
of the main text can be written as
\begin{equation}
H = H_0 + \sum_j \sum_{\langle \mathbf{r}, \tilde{\mathbf{r}}
\rangle_j} V_{\langle \mathbf{r}, \tilde{\mathbf{r}} \rangle_j},
\qquad H_0 = - \lambda J \sum_{\mathbf{r}} \sum_{\langle j,j'
\rangle} \sigma_{\mathbf{r}, j}^z \sigma_{\mathbf{r}, j'}^z, \qquad
V_{\langle \mathbf{r}, \tilde{\mathbf{r}} \rangle_j} = -J
\sigma_{\mathbf{r}, j}^x \sigma_{\tilde{\mathbf{r}}, j}^y,
\label{eq-pert-H-1}
\end{equation}
where the first term $H_0$ is the unperturbed Hamiltonian, and
$V_{\langle \mathbf{r}, \tilde{\mathbf{r}} \rangle_j}$ in the second
term are the perturbations. For a lattice of $N$ sites, the
ground-state subspace of $H_0$ contains $2^N$ degenerate states $|
\Psi_0 \rangle$ with energies $E_0 = -6N \lambda J$. At order $p$ in
degenerate (Brillouin-Wigner) perturbation theory, the low-energy
Hamiltonian induced within this ground-state subspace is then
\begin{equation}
\tilde{H}_p = \sum_{j_1, \ldots, j_p} \sum_{\langle \mathbf{r}_1,
\tilde{\mathbf{r}}_1 \rangle_{j_1}} \ldots \sum_{\langle
\mathbf{r}_p, \tilde{\mathbf{r}}_p \rangle_{j_p}} \mathcal{P}
\prod_{l=1}^{p-1} \left[ V_{\langle \mathbf{r}_l,
\tilde{\mathbf{r}}_l \rangle_{j_l}} \left( E - H_0 \right)^{-1}
\left( 1 - \mathcal{P} \right) \right] V_{\langle \mathbf{r}_p,
\tilde{\mathbf{r}}_p \rangle_{j_p}} \mathcal{P}, \label{eq-pert-H-2}
\end{equation}
where the projector $\mathcal{P} = \prod_{\mathbf{r}} \prod_{\langle
j,j' \rangle} (1 + \sigma_{\mathbf{r}, j}^z \sigma_{\mathbf{r},
j'}^z)$ annihilates any state outside the ground-state subspace, and
the energy $E$ can (in principle) be determined self-consistently
via $E = E_0 + \sum_{q = 1}^{p} \langle \Psi_0 | \tilde{H}_q |
\Psi_0 \rangle$. Importantly, the unperturbed Hamiltonian $H_0$ is
an exclusive function of commuting $\mathbb{Z}_2$ operators
$\sigma_{\mathbf{r}, j}^z \sigma_{\mathbf{r}, j'}^z = \pm 1$, and
any state $| \Psi_0 \rangle$ in its ground-state subspace is
characterized by $\sigma_{\mathbf{r}, j}^z \sigma_{\mathbf{r}, j'}^z
= +1$ for all $\mathbf{r}$, $j$, and $j'$. Since each perturbation
term $V_{\langle \mathbf{r}, \tilde{\mathbf{r}} \rangle_j}$ either
commutes or anticommutes with each $\sigma_{\mathbf{r}, j}^z
\sigma_{\mathbf{r}, j'}^z$, each resolvent $(E - H_0)^{-1}$ in
Eq.~(\ref{eq-pert-H-2}) acts on an eigenstate of $H_0$ characterized
by the same definite values $\sigma_{\mathbf{r}, j}^z
\sigma_{\mathbf{r}, j'}^z = \pm 1$ for any state $| \Psi_0 \rangle$
in the ground-state subspace acted upon by $\tilde{H}_p$.
Substituting the resolvents with their corresponding eigenvalues,
the low-energy Hamiltonian in Eq.~(\ref{eq-pert-H-2}) then takes the
form
\begin{equation}
\tilde{H}_p = \sum_{j_1, \ldots, j_p} \sum_{\langle \mathbf{r}_1,
\tilde{\mathbf{r}}_1 \rangle_{j_1}} \ldots \sum_{\langle
\mathbf{r}_p, \tilde{\mathbf{r}}_p \rangle_{j_p}} \Lambda_{\langle
\mathbf{r}_1, \tilde{\mathbf{r}}_1 \rangle_{j_1}, \ldots, \langle
\mathbf{r}_p, \tilde{\mathbf{r}}_p \rangle_{j_p}} \mathcal{P}
\prod_{l=1}^{p-1} \left[ V_{\langle \mathbf{r}_l,
\tilde{\mathbf{r}}_l \rangle_{j_l}} \left( 1 - \mathcal{P} \right)
\right] V_{\langle \mathbf{r}_p, \tilde{\mathbf{r}}_p \rangle_{j_p}}
\mathcal{P}, \label{eq-pert-H-3}
\end{equation}
where each $\Lambda_{\langle \mathbf{r}_1, \tilde{\mathbf{r}}_1
\rangle_{j_1}, \ldots, \langle \mathbf{r}_p, \tilde{\mathbf{r}}_p
\rangle_{j_p}} \sim 1 / (\lambda J)^{p-1}$ is a product of resolvent
eigenvalues. Due to the projectors $\mathcal{P}$ and $1 -
\mathcal{P}$, the only non-vanishing terms in $\tilde{H}_p$ are the
ones where (i) each intermediate state in between the perturbation
terms $V_{\langle \mathbf{r}_l, \tilde{\mathbf{r}}_l \rangle_{j_l}}$
is outside the ground-state subspace and (ii) the product
$\prod_{l=1}^p V_{\langle \mathbf{r}_l, \tilde{\mathbf{r}}_l
\rangle_{j_l}}$ can be expressed entirely in terms of the effective
spin operators $X_{\mathbf{r}}$, $Y_{\mathbf{r}}$, and
$Z_{\mathbf{r}}$ in Eq.~(2) of the main text. It can then be checked
with the aid of a computer that the lowest-order non-trivial (i.e.,
non-constant) terms are obtained at order $p = 32$ in perturbation
theory and that the resulting low-energy Hamiltonian is
$\tilde{H}_{32} = \sum_{\mathbf{r}} W_{\mathbf{r}}$, where
$W_{\mathbf{r}}$ is given in Eq.~(3) of the main text. Furthermore,
it can be verified that any terms obtained at higher orders of
perturbation theory are products of $W_{\mathbf{r}}$. Note, however,
that the coefficient of $W_{\mathbf{r}}$ in Eq.~(3) of the main text
is not straightforward to evaluate as it is the sum of contributions
from $32!$ different coefficients $\Lambda_{\langle \mathbf{r}_1,
\tilde{\mathbf{r}}_1 \rangle_{j_1}, \ldots, \langle \mathbf{r}_{32},
\tilde{\mathbf{r}}_{32} \rangle_{j_{32}}}$.

\section{Relation to coupled-layer constructions} \label{sec-lay}

Here we demonstrate that the low-energy Hamiltonian $\tilde{H}$ can
be obtained from the coupled-spin-chain Hamiltonian $H$ in two
consecutive steps via an intermediate coupled-layer Hamiltonian. In
the first step, we form two pairs $j = 1,2$ and $j = 3,4$ out of the
four spin flavors $j = 1,2,3,4$ in Eq.~(1) of the main text and
restrict the on-site couplings to act only within the individual
pairs. The coupled-spin-chain Hamiltonian takes the modified form
\begin{equation}
H' = -J \sum_j \sum_{\langle \mathbf{r}, \tilde{\mathbf{r}}
\rangle_j} \sigma_{\mathbf{r}, j}^x \sigma_{\tilde{\mathbf{r}}, j}^y
- \lambda J \sum_{\mathbf{r}} \left( \sigma_{\mathbf{r}, 1}^z
\sigma_{\mathbf{r}, 2}^z + \sigma_{\mathbf{r}, 3}^z
\sigma_{\mathbf{r}, 4}^z \right) = \sum_{(1 \, 0 \, \bar{1})} H_{(1
\, 0 \, \bar{1})} + \sum_{(1 \, 0 \, 1)} H_{(1 \, 0 \, 1)},
\label{eq-lay-H-1}
\end{equation}
where the spin flavors $j = 1,2$ and $j = 3,4$ are only coupled
within individual $(1 \, 0 \, \bar{1})$ and $(1 \, 0 \, 1)$ planes
by the Hamiltonians
\begin{eqnarray}
H_{(1 \, 0 \, \bar{1})} &=& -J \sum_{\mathbf{r} \in (1 \, 0 \,
\bar{1})} \left[ \sigma_{\mathbf{r}, 1}^x \sigma_{\mathbf{r} +
\mathbf{b}_1, 1}^y + \sigma_{\mathbf{r}, 2}^x \sigma_{\mathbf{r} +
\mathbf{b}_2, 2}^y \right] - \lambda J \sum_{\mathbf{r} \in (1 \, 0
\, \bar{1})} \sigma_{\mathbf{r}, 1}^z \sigma_{\mathbf{r}, 2}^z,
\nonumber \\
H_{(1 \, 0 \, 1)} &=& -J \sum_{\mathbf{r} \in (1 \, 0 \, 1)} \left[
\sigma_{\mathbf{r}, 3}^x \sigma_{\mathbf{r} + \mathbf{b}_3, 3}^y +
\sigma_{\mathbf{r}, 4}^x \sigma_{\mathbf{r} + \mathbf{b}_4, 4}^y
\right] - \lambda J \sum_{\mathbf{r} \in (1 \, 0 \, 1)}
\sigma_{\mathbf{r}, 3}^z \sigma_{\mathbf{r}, 4}^z.
\label{eq-lay-H-2}
\end{eqnarray}
In the strong-coupling regime ($\lambda \gg 1$), the two spins
$\sigma_{\mathbf{r}, j}$ within each pair $j = 1,2$ and $j = 3,4$
are locked together by the on-site couplings at each site
$\mathbf{r}$. The local Hilbert space is then captured by two
effective spins $\mu_{\mathbf{r}}$ and $\tau_{\mathbf{r}}$ as its
four states can be characterized by $\mu_{\mathbf{r}}^z =
\sigma_{\mathbf{r}, 1}^z = \sigma_{\mathbf{r}, 2}^z = \pm 1$ and
$\tau_{\mathbf{r}}^z = \sigma_{\mathbf{r}, 3}^z =
\sigma_{\mathbf{r}, 4}^z = \pm 1$. For each coupled-spin-chain
Hamiltonian in Eq.~(\ref{eq-lay-H-2}), we employ degenerate
perturbation theory to obtain a low-energy Hamiltonian in terms of
these effective spins. Treating the second (on-site) term as the
unperturbed Hamiltonian and the first (nearest-neighbor) term as the
perturbation, the lowest-order non-trivial Hamiltonian terms arise
at order $4$ in perturbation theory, and the resulting low-energy
Hamiltonians are given by
\begin{eqnarray}
\tilde{H}_{(1 \, 0 \, \bar{1})} &\sim& \frac{J} {\lambda^{3}}
\sum_{\mathbf{r} \in (1 \, 0 \, \bar{1})} \mu_{\mathbf{r}}^x \,
\mu_{\mathbf{r} + \mathbf{b}_1}^y \, \mu_{\mathbf{r} +
\mathbf{b}_2}^y \, \mu_{\mathbf{r} + \mathbf{b}_1 + \mathbf{b}_2}^x,
\nonumber \\
\tilde{H}_{(1 \, 0 \, 1)} &\sim& \frac{J} {\lambda^{3}}
\sum_{\mathbf{r} \in (1 \, 0 \, 1)} \tau_{\mathbf{r}}^x \,
\tau_{\mathbf{r} + \mathbf{b}_3}^y \, \tau_{\mathbf{r} +
\mathbf{b}_4}^y \, \tau_{\mathbf{r} + \mathbf{b}_3 +
\mathbf{b}_4}^x. \label{eq-lay-H-3}
\end{eqnarray}
Importantly, the two-dimensional Hamiltonians $\tilde{H}_{(1 \, 0 \,
\bar{1})}$ and $\tilde{H}_{(1 \, 0 \, 1)}$ are topologically ordered
as they are each equivalent to the toric-code Hamiltonian up to
canonical transformations (see Refs.~[1] and [23] in the main text).

In the second step, we restore the remaining on-site couplings
between spin flavors $j = 1,2$ and $j = 3,4$. In terms of the
effective spins $\mu_{\mathbf{r}}$ and $\tau_{\mathbf{r}}$ and the
single-layer Hamiltonians in Eq.~(\ref{eq-lay-H-3}), the Hamiltonian
then takes the coupled-layer form
\begin{equation}
\hat{H} = \sum_{(1 \, 0 \, \bar{1})} \tilde{H}_{(1 \, 0 \, \bar{1})}
+ \sum_{(1 \, 0 \, 1)} \tilde{H}_{(1 \, 0 \, 1)} - 4 \lambda J
\sum_{\mathbf{r}} \mu_{\mathbf{r}}^z \tau_{\mathbf{r}}^z.
\label{eq-lay-H-4}
\end{equation}
In the strong-coupling regime ($\lambda \gg 1$), the two effective
spins $\mu_{\mathbf{r}}$ and $\tau_{\mathbf{r}}$ at each site
$\mathbf{r}$ are locked together by the remaining on-site couplings.
The local Hilbert space is then captured by a single effective spin
$\Sigma_{\mathbf{r}}$ as its two states can be characterized by
$\Sigma_{\mathbf{r}}^z = \mu_{\mathbf{r}}^z = \tau_{\mathbf{r}}^z =
\sigma_{\mathbf{r}, j}^z = \pm 1$. To obtain a low-energy
Hamiltonian in terms of these effective spins, we employ degenerate
perturbation theory, treating the third (on-site) term as the
unperturbed Hamiltonian and the first two (single-layer) terms as
the perturbations. The lowest-order non-trivial Hamiltonian terms
arise at order $8$ in perturbation theory, and the resulting
low-energy Hamiltonian is $\tilde{H} = \sum_{\mathbf{r}}
W_{\mathbf{r}}$, where $W_{\mathbf{r}}$ is given in Eq.~(3) of the
main text. Remarkably, the coupled-layer Hamiltonian in
Eq.~(\ref{eq-lay-H-4}) is analogous to the coupled-layer
constructions introduced in Refs.~[17] and [18] of the main text.
Indeed, the fracton topological order is obtained in our second step
by strongly coupling orthogonal stacks of two-dimensional
topologically ordered layers. Note, however, that we have only two
(rather than three) such orthogonal stacks (see also Ref.~[19] in
the main text).

\clearpage

\end{widetext}



\begin{references}
\bibitem{Footnote-1} In fact, for an $L \times L \times L$ system, a detailed
analysis shows that the ground-state degeneracy is $2^{12L-10}$.
\bibitem{Footnote-2} The coupled-spin-chain models in this work
typically describe type-I fracton phases as they have excitations
that are mobile along the individual spin chains. Nevertheless,
type-II fracton phases with only immobile excitations can in
principle also be captured if all spin chains form closed loops.
This extension of our work is currently in progress.
\end{references}
\end{document}